\documentclass[twocolumn,showpacs,preprintnumbers,amsmath,amssymb]{revtex4}
\usepackage{epsfig,amsfonts}
\usepackage{amsmath}
\usepackage{bbm}
\usepackage{graphicx,psfrag,rotating}
\usepackage{graphics}
\usepackage{txfonts}
\usepackage{graphicx}% Include figure files
\usepackage{dcolumn}% Align table columns on decimal point
\usepackage{bm}% bold math

\begin{document}

%\preprint{APS/123-QED}%

\title{ Quantum Optomechanics in the Bistable Regime }

\author{R. Ghobadi$^1,^2$,  A.R. Bahrampour$^2$, and  C. Simon$^1$}
\affiliation{$^1$ Institute for Quantum Information Science
and Department of Physics and Astronomy, University of
Calgary, Calgary T2N 1N4, Alberta, Canada\\$^2$Department of Physics, Sharif University of Technology, Tehran, Iran}

\date{\today}

\begin{abstract}
We study the simplest optomechanical system with a focus on the bistable regime.
The covariance matrix formalism allows us to study both cooling and entanglement in a unified framework. We identify two key factors governing entanglement, namely the bistability parameter, i.e. the distance from the end of a stable branch in the bistable regime, and the effective detuning, and we describe the optimum regime where entanglement is greatest. We also show that in general entanglement is a non-monotonic function of optomechanical coupling. This is especially important in understanding the optomechanical entanglement of the second stable branch.
\end{abstract}

\pacs{}
\maketitle
\section{\label{intro}Introduction }

Observing quantum effects like superposition states or entanglement at the macroscopic level is a long standing goal. It is a widely held view that this should be possible, provided that enviromentally induced decoherence can be sufficiently suppressed. Note however that there are some theoretical proposals which would rule out the existence of quantum effects at the macroscopic level, see e.g. Ref. \cite{Penrose00}. Proposals for the experimental observation of macroscopic quantum effects are often based on the principle of Schr\"{o}dinger's cat, i.e. on coupling a microscopic quantum system to a macroscopic system in a controlled way, in order to create a macroscopic superposition state \cite{Armour02,Bose99,Marshall03}.

One particularly promising approach in this context is the use of optomechanical systems. The most basic optomechanical system consists of a Fabry-Perot
cavity with one movable end mirror. The position of this mirror is determined by the radiation pressure inside the cavity. Such systems were first studied in the context
 of high precision measurements and gravitational wave detection \cite{Braginsky77}. It was suggested in Ref. \cite{Bose99} that
 the radiation pressure of a single photon in a high finesse optical cavity could in principle create a macroscopic superposition of two spatially
 distinct locations of a moveable mirror . A potential implementation of this idea was proposed in Ref.  \cite{Marshall03}. It is very challenging experimentally to achieve sufficiently strong optomechanical coupling at the single-photon level, requiring a system that combines high optical and mechanical finesse, low mechanical resonance frequency and ultra-low temperature.

One way to enhance the opto-mechanical interaction is to pump the cavity with a strong laser. Using this technique the strong coupling regime in optomechanical systems has recently been reached\cite{Groblacher09}.
In the presence of a strong enough driving laser the field enhancement inside the high finesse optical cavity is large enough to trigger
nonlinear behaviour of the system. Depending on the input power and the detuning of the driving laser with respect to the cavity resonance, optomechanical systems exhibit different types of nonlinear behaviour. For strong enough input power, in the blue detuned regime one obtains multistability\cite{Marquardt06}, instability\cite{O. Arcizet06} and chaotic motion \cite{Carmon07}. In the red detuned regime bistability \cite{Dorsel83,Karuza} occurs. Here we consider the red detuned regime. This is the appropriate regime for cooling the mechanical oscillator close to the ground state \cite{Riviere,Teufel}, which is usually seen as a prerequisite for observing quantum effects. We are particularly interested in the relationship between bistability and entanglement.

Optomechanical bistability can be understood intuitively as the result of a competition between the mechanical restoring force, which increases linearly when the mirror is moved from its equilibrium position, and the radiation pressure force, which has a maximum at the cavity resonance. For a suitable set of  parameters as in Fig.1  there are three intersection points between the two forces. The leftmost and rightmost intersection points correspond to stable states, because the restoring force grows faster than the radiation pressure (for the rightmost point the radiation pressure even decreases as the mirror is pushed outwards). In contrast, the middle intersection point is unstable because the radiation pressure force increases faster than the restoring force.

\begin{figure}
\epsfig{file=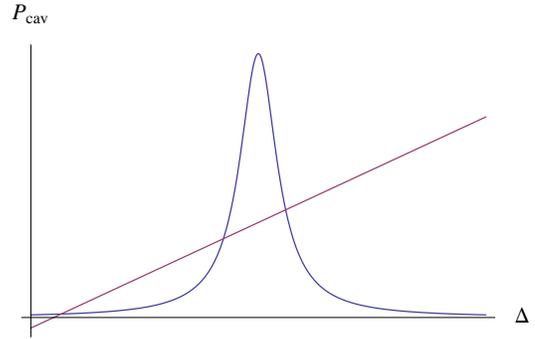,width=0.8 \columnwidth}
%\scalebox{1}{\includegraphics*[viewport=-10 0 500 85]{1.eps}}
\caption{\label{fig1} Mechanical restoring force and radiation pressure force around a cavity resonance. The leftmost and rightmost intersection points are stable equilibrium positions, whereas the middle one is unstable.}
\end{figure}

It is known that the optomechanical interaction can squeeze the cavity mode, and this squeezing becomes maximal close to the bistable regime \cite{Fabre94}.
It has also been noted\cite{C.Genes08} that under certain conditions entanglement is maximized at the bistability threshold. This was interpreted as being due to the enhanced optomechanical coupling strength in this region. Here we analyze entanglement close to and in the bistable regime in detail. We show that a lot of insight can be gained by analyzing the situation in terms of two key parameters, namely the effective detuning and the bistability parameter, which quantifies the distance from the end of each bistable branch. Cooling and entanglement can be studied in the same theoretical framework based on the covariance matrix. We identify the optimal regimes for both cooling and entanglement. We also show that, somewhat surprisingly, entanglement is in general a non-monotonic function of the optomechanical coupling strength. (Naively one might have expected it to always increase with optomechanical coupling strength .)

The paper is organized as follows: Section II introduces the optomechanical system and describes the linearization of the equations of motion around the steady state. We also show how bistability arises in the red-detuned regime in this framework, introduce the bistability parameter, and derive the dependence of the photon and phonon number on this parameter, which leads us to a discussion of cooling.
Section III discusses the optomechanical entanglement and its dependence on the bistability parameter and the effective detuning. This allows us to
determine the optimum value for the detuning and the maximum achievable entanglement in our system. We discuss the role of the optomechanical coupling constant, show how entanglement varies on both stable branches in the bistable regime, and discuss its robustness under increasing temperature.
Section IV is a summary and conclusion.

\section{\label{ System} The System}

We consider a high Q Fabry-Perot cavity with decay rate $\kappa$.
The moveable mirror can move under the influence of
radiation pressure and thermal noise. The moveable mirror is initially in equilibrium
with a thermal bath at temperature $T$ which results in the mechanical damping
rate $\gamma_{m}$ and the noise force $\xi(t)$. The system is driven
by a laser with frequency $\omega_{L}$ and power $P$. The general
Hamiltonian of such a system is derived in \cite{Law95}. In the regime of
parameters that we are interested in, the general Hamiltonian simplifies to
\cite{Law95,Vitali07}
\begin{equation}
H=\hbar\omega_{c}a^{+}a+\frac{\hbar\omega_{m}}{2}(p^{2}+q^{2})-\hbar G_{0}a^{+}aq+i\hbar E(a^{+}e^{-i\omega_{L}t}-ae^{i\omega_{L}t}),\end{equation}
where $\omega_{c}$ and $a$ are frequency and annihilation operator
of the cavity mode,respectively, $\omega_{m}$, $q$,$p$ are frequency
and dimensionless position and momentum operator of the mirror,respectively,
and $G_{0}=\frac{\omega_{c}}{L}\sqrt{\frac{\hbar}{m\omega_{m}}}$
is the coupling constant and $E=\sqrt{\frac{2P\kappa}{\hbar\omega_{L}}}$
where $P$, $\omega_{L}$ are the input laser power and frequency
respectively.
The first two terms correspond to two free harmonic oscillators,
the third term corresponds to the optomechanical coupling and the last
term corresponds to the cavity being driven by the laser.

The equations of motion in the presence of damping and noise are
\begin{equation}
\dot{q}=\omega_{m}p\end{equation}
\begin{equation}
\dot{p}=-\omega_{m}q-\gamma_{m}p+G_{0}a^{+}a+\xi(t)\end{equation}
\begin{equation}
\dot{a}=-(\kappa+i\Delta_{0})a+iG_{0}aq+E+\sqrt{2\kappa}a_{in}\end{equation}
where $\Delta_{0}=\omega_{c}-\omega_{L}$, $a_{in}$ is the vacuum input noise of the cavity, and $\xi(t)$ is the noise associated with the damping of the mechanical oscillator.
The nonlinear Eqs. (3,4) can be linearized by expanding the operators around their steady state values $O_{i}=O_{i,s}+\delta O_{i}$ ,where $O_{i}=a,q,p$.

\begin{figure}
\epsfig{file=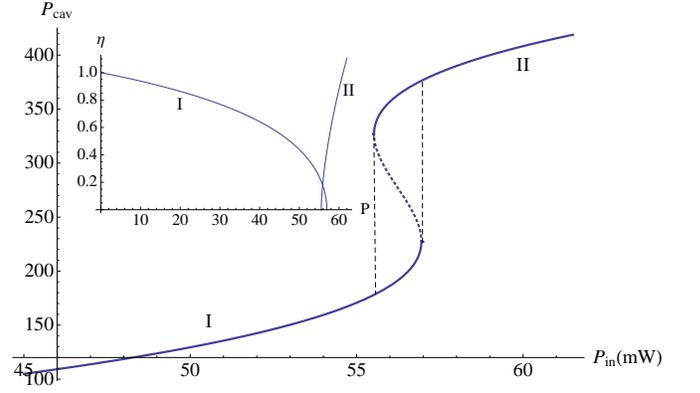,width=\columnwidth}
%\scalebox{1.1}{\includegraphics*[viewport=5 0 320 130]{Figure2.eps}}
\caption{\label{fig3} Bistability of the intracavity power with respect to the input power. The solid and dotted lines correspond to the stable and unstable branches respectively. The inset shows the bistability parameter $\eta$ for the two stable branches. The end of each stable branch corresponds to $\eta=0$.
}
\end{figure}

From Eqs. (2,3,4), the steady state solutions are
$\alpha_{s}=\frac{E}{\kappa+i(\Delta_{0}-G_{0}q_{s})}$, $q_{s}=\frac{G_{0}\mid\alpha_{s}\mid^{2}}{\omega_{m}}$, $p_{s}=0$,
 where $\alpha_{s}$,$q_{s}$,$p_{s}$are the stationary values for cavity amplitude, position and momentum of mechanical oscillator, respectively. Note that the last of these relations is a third order polynomial equation for $\alpha_s$, which has three roots. The largest and the smallest roots are stable, and the middle one is unstable.
Fig. 2 shows the hysteresis loop for the intracavity power. Consider $P_{cav}$  initially on the lower stable branch (I in Fig. 2, corresponding to the smallest root). As $P_{in}$ increases past its value for the first turning point, $P_{cav}$ switches to the upper stable branch (II in Fig. 2, corresponding to the largest root). For  $P_{in}$  larger than its value at this switch point , $P_{cav}$ is given by the upper branch. As $P_{in}$ decreases below this value $P_{cav}$ is still given by the upper branch until $P_{in}$ decreases below its value for the upper turning point. At this point $P_{cav}$ switches back down to the lower branch. We have used the the set of parameters of Ref\cite{Vitali07}, which is close to several optomechanical experiments\cite{Kleckner06,Gigan06,DKleckner06,Carmon05}.
 We consider Fabry-Perot cavity  with length $L=1mm$ and finesse $\mathcal{F}=1.07\times10^{4}$, driven by a laser with $\lambda=810 nm$ and $\Delta_{0}=2.62\omega_{m}$. The mechanical oscillator frequency ,damping rate and mass are $10MHz$,$100Hz$ and $5ng$ respectively with environment temperature $T=400mK$.

By introducing $u^{T}(t)=(\delta q(t),\delta p(t),X(t),Y(t))$ and $n^{T}(t)=(0,\xi(t),\sqrt{2\kappa}X_{in}(t),\sqrt{2\kappa}Y_{in}(t))$ where $X=\frac{\delta a+\delta a^{+}}{\sqrt{2}}$ and $Y=\frac{\delta a-\delta a^{+}}{\sqrt{2}i}$ and corresponding noises $X_{in}$and $Y_{in}$, the linearized dynamics of system can be written in a compact form
 \begin{equation}
\dot{u}(t)=Au(t)+n(t)\end{equation}
 where
\begin{equation}
A=\left(\begin{array}{cccc}
0 & \omega_{m} & 0 & 0\\
-\omega_{m} & -\gamma_{m} & G & 0\\
0 & 0 & -\kappa & \Delta\\
G & 0 & -\Delta & -\kappa\end{array}\right)\end{equation}
 and $G=\sqrt{2}G_{0}\alpha_{s}$ , $\Delta=\Delta_{0}-G_{0}q_{s}$
are the enhanced optomechanical coupling rate and effective detuning.

Since the initial state of the system is Gaussian and the dynamical
equations are linear in the creation and annihilation operators
both for cavity and mechanical mode the state of the system remains
Gaussian at all times. A Gaussian state is fully characterized by its
covariance matrix which is defined at any given time $t$ by $V_{ij}(t)=\frac{\langle u_{i}(t)u_{j}(t)+u_{j}(t)u_{i}(t)\rangle}{2}$.
The mechanical and the optical input noise operators are fully characterized by their correlation function which in the Markovian approximation are given by
\begin{equation}
\langle a_{in}(t)a_{in}^{+}(t')\rangle=\delta(t-t').\end{equation}
\begin{equation}
\frac{\langle\xi(t)\xi(t')+\xi(t')\xi(t)\rangle}{2}=\gamma_{m}(2\bar{n}+1)\delta(t-t').\end{equation}
where $\overline{n}=[exp(\frac{\hbar\omega_{m}}{k_{B}T})-1]^{-1}$
is the mean thermal phonon number and $k_{B}$ is Boltzmann's constant.
From Eqs.(5,7,8) one obtains the equation of motion for the covariance
matrix which is given by \cite{Mari09} \begin{equation}
\dot{V}=AV+VA^{T}+D.\end{equation}
The steady state solution for the covariance matrix ($\dot{V}=0$) is reached if all the eigenvalues of the matrix $A$ have negative real parts. In the red detuned regime of operation ($\Delta>0$), the Routh-Hurwitz criterion \cite{Dejesus87} gives the following stability condition
\begin{equation}
\omega_{m}(\kappa^{2}+\Delta^{2})-G^{2}\Delta>0.
\label{stabilitycond}
\end{equation}
In the following we use the dimensionless "bistability parameter" defined as \cite{Genes08}
\begin{equation}
\eta=1-\frac{G^{2}\Delta}{\omega_{m}(\kappa^{2}+\Delta^{2})}
\end{equation}
which is a positive number between zero and one according to Eq.(\ref{stabilitycond}) in the red detuned regime ($\Delta>0$). We have shown the bistability parameter in the inset of Fig.2. As can be seen from Fig.2, $\eta$ decreases when approaching the bistable regime and becomes equal to zero at the end of each stable branch.

Eq. (10) can be intuitively understood by ignoring retardation effects for the radiation pressure. Assuming that the optical field adiabatically follows the mechanical oscillator (i.e. setting $\dot{\delta a}=0$), one has $\delta a=\frac{1}{(\kappa+i\Delta)}(\frac{iG}{\sqrt{2}}\delta q+\sqrt{2\kappa}a_{in})$.
 The equation of motion for the mirror becomes
\begin{equation}
\dot{\delta p}=-(\omega_{m}-\frac{G^{2}\Delta}{\kappa^{2}+\Delta^{2}})\delta q-\gamma_{m}\delta p+\xi_{T} \label{deltapdot} \end{equation}
where $\xi_{T}=\xi+G\sqrt{2\kappa}(\frac{a_{in}}{\kappa+i\Delta}+\frac{a_{in}^{+}}{\kappa-i\Delta})$.
From Eq.(11), we see that the mechanical oscillator is stable if the first coefficient is positive. In this case the first term in Eq. (\ref{deltapdot})  corresponds to a harmonic restoring force, see also Fig. 1 and the associated discussion. This implies the stability condition (10). The adiabatic approximation is equivalent to treating the response of the cavity field to the moving mirror as instantaneous. It is well known that the delayed nature of this response gives rise to cooling\cite{Karrai04}, which is however not essential for the above argument. We feel that this argument helps the physical understanding of the stability condition. However, let us emphasize that we will not make the adiabatic approximation in what follows.

In the bistable regime the fluctuations around the steady state solution diverge as one approaches the end of each stable branch. To show this explicitly we solve the Eq.(9) for steady state, from which we can obtain the phonon and photon numbers by using $\bar{n}_{m}=\frac{ V_{11}+V_{22}-1}{2}$
and $\bar{n}_{o}=\frac{ V_{33}+V_{44}-1}{2}$. The general solution is complicated and not very illuminating. Simple relations that show the dependence of the fluctuations on the stability parameter can be obtained by assuming a high mechanical quality factor and low temperature environment,i.e. $\frac{\omega_{m}}{\gamma_{m}}>>1$
and $\frac{\kappa}{\bar{n}\gamma_{m}}>>1$ . We find
\begin{equation}
\bar{n}_{m}=\frac{(\Delta^{2}+\kappa^{2})(1+\eta)-2\eta\omega_{m}(2\Delta-\omega_{m})}{8\Delta\eta\omega_{m}}.\end{equation}
\begin{equation}
\bar{n}_{o}=\frac{(1-\eta)(\kappa^{2}+\Delta^{2})}{8\eta\Delta^{2}}.\end{equation}
From Eq.{(13,14)} it is clear that the phonon and photon numbers diverge as $\eta$ approaches zero. In order to stay within the range of validity of the linearization approximation, we have made sure that $\bar{n}_o \ll {\mid\alpha_{s}\mid^{2}}$ in all the results shown below.

For $\eta\sim1$ from Eq.(13) the optimum value for the detuning which minimizes the phonon number is given by
\begin{equation}
 \Delta_{opt}=\sqrt{\kappa^{2}+\omega_{m}^{2}}\end{equation}
 Using the optimum value for detuning in Eq.(13) one finds
\begin{equation}
\bar{n}_{m}=\frac{1}{2}(\frac{\sqrt{\kappa^{2}+\omega_{m}^{2}}}{\omega_{m}}-1)\end{equation}
which is identical to Eq.(7) in \cite{Wilson-Rae07} . In the resolved sideband regime ($\omega_{m}>>\kappa$) one sees from Eq. (15) that ground state cooling can be achieved
($\bar{n}_{m}=\frac{\kappa^{2}}{4\omega_{m}^{2}}$)\cite{Marquardt07,Wilson-Rae07}.

\section{\label{Entanglement}Opto-Mechanical entanglement}

As shown in \cite{Simon00}, for bipartite Gaussian states the Peres- Horodecki criterion\cite{Peres96,Horodecki97} (positivity of the density matrix under partial transposition) is necessary and sufficient for separability. In terms of the covariance matrix formalism this criterion is called logarithmic negativity which is defined as \cite{Adesso04}
\begin{equation}
E_{N}=max\{0,-ln(2\nu_{min})\}.\end{equation}
where $\nu_{min}$ is the smallest symplectic eigenvalue of the partially
transposed covariance matrix given by $\nu_{min}=\sqrt{\frac{\Sigma-\sqrt{\Sigma^{2}-4detV}}{2}}$, where $\Sigma=detA+detB-2detC$, and we represent the covariance matrix in terms of
\begin{equation}
V=\left(\begin{array}{cc}
A & C\\
C^{T} & B\end{array}\right).\end{equation}

\begin{figure}
\scalebox{1.4}{\includegraphics*[viewport=0 0 200 290]{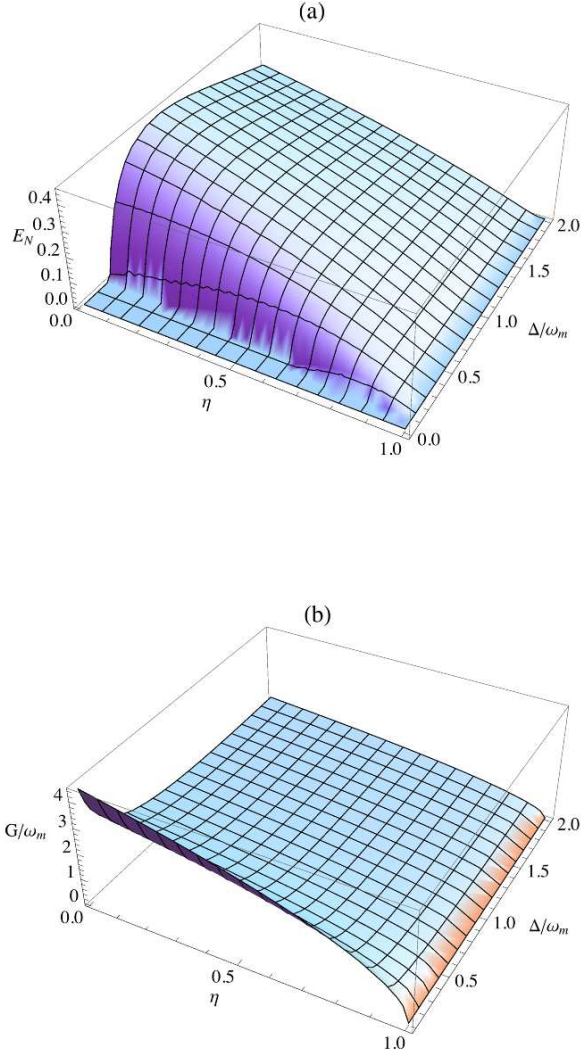}}
\caption{\label{fig3} Optomechanical entanglement (a) and optomechanical coupling constant (b) as a function of bistability parameter $\eta$ and normalized effective detuning $\frac{\Delta}{\omega_{m}}$, for a cavity decay rate $\kappa=1.4$$\omega_{m}$.}
\end{figure}

Equipped with this measure we go on to study optomechanical entanglement. Fig. 3(a) shows the logarithmic negativity as a function of bistability parameter $\eta$ and effective detuning $\Delta$. We note that for $\eta\sim1$, which is required for ground state cooling, there is no optomechanical entanglement.
From Figure 3(a) one can identify three different regimes depending on the effective detuning. In the first regime, corresponding to $\Delta<0.1\omega_{m}$ in the figure, there is no optomechanical entanglement.
In the second regime, corresponding to $0.1\omega_{m}<\Delta<0.3\omega_{m}$ in the figure, there is some optomechanical entanglement but the maximum value for entanglement is attained for values of the bistability parameter $\eta$ somewhere between zero and one. This shows that for a fixed detuning, the maximum entanglement does not necessarily occur at the end of the bistable branch, cf. \cite{C.Genes08}.
Finally in the third regime, $\Delta>0.3\omega_{m}$ in the figure, there is strong optomechanical entanglement, and for each fixed value of detuning the maximum entanglement is in fact reached at the end of the branch, i.e. for $\eta=0$. So in short we have $E_{N,1}<E_{N,2}<E_{N,3}$,where $E_{N,i}$ is logarithmic negativity in the $i$-th regime.

Fig.3(b) shows the corresponding optomechanical coupling in the different regimes. From Fig.3(b) it is clear that optomechanical coupling is monotonically decreasing   function of effective detuning i.e  $G_{1}>G_{2}>G_{3}$ where $G_{i}$ is the optomechanical coupling constant in the $i$-th regime. So we see that in general entanglement is not a monotonically increasing function of the coupling constant.
These observations suggest that the key variables that determine the entanglement behaviour are the effective detuning and the bistability parameter, not the optomechanical coupling constant.

 A more quantitative understanding of the different regimes for entanglement is possible by looking at the entanglement behaviour in the vicinity of $\eta=0$.
 Assuming that $\frac{\omega_{m}}{\gamma_{m}}>>1$
and $\frac{\kappa}{\bar{n}\gamma_{m}}>>1$ one finds $\Sigma=a+\frac{b}{\eta}$ and
$detV=c+\frac{d}{\eta}$, where
\begin{equation}
a=\frac{\Delta^{2}-3\kappa^{2}+\omega_{m}^{2}}{16\Delta^{2}}\end{equation}

\begin{equation}
b=\frac{(\Delta^{2}+\kappa^{2})(\Delta^{2}+\kappa^{2}+5\omega_{m}^{2})}{16\Delta^{2}\omega_{m}^{2}}\end{equation}

\begin{equation}
c=\frac{2\Delta^{2}(\Delta^{2}+\kappa^{2})+(\Delta^{2}-\kappa^{2})\omega_{m}^{2}}{128\Delta^{4}}\end{equation}

\begin{equation}
d=\frac{(\Delta^{2}+\kappa^{2})(4\Delta^{4}+4\Delta^{2}\kappa^{2}+4\Delta^{2}\omega_{m}^{2}+\omega_{m}^{4})}{256\Delta^{4}\omega_{m}^{2}}\end{equation}
From these equations it is possible to drive a simple form for logarithmic negativity.
Close to the bistability region ($\eta<<1$) we have $E_{N}=max\{0,\alpha+\beta\eta\}$ where $\alpha=-ln(2\sqrt{\frac{d}{b}})$ and $\beta=\frac{(abd-b^{2}c-d^{2})}{2db^{2}}$.

It is worth noting that in contrast to the phonon and photon numbers, which diverge for $\eta=0$, the logarithmic negativity has a finite limiting value given by $\alpha$. While our linearization approximation is not justified for the point $\eta=0$ itself, it does apply in its close vicinity, as the photon number drops precipitously as one moves away from the end point of the stable branch, cf. Eq. (14).

Using our expression for entanglement close to bistability one can easily identify the three regimes shown in Fig.3. The first regime corresponds to $\alpha,\beta<0$. The second regime corresponds to $\alpha<0,\beta>0$ or $\alpha,\beta>0$ and the third region corresponds to $\alpha>0,\beta<0$.

\begin{figure}
\scalebox{0.55}{\includegraphics*[viewport=0 0 500 280]{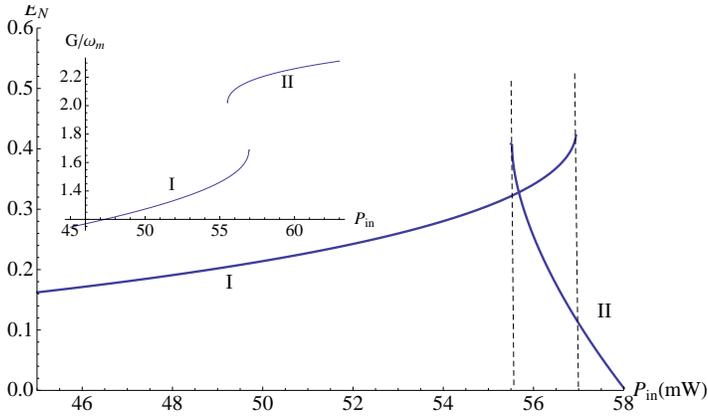}}
\caption{\label{fig1}Plot of the optomechanical entanglement as a function of input power for both stable branches.
The dot dashed (dashed) line corresponds to the end of the first (second) stable branch. The parameters are the same as in Fig.2.}
\end{figure}

Moreover as can be seen from Fig. 3 the maximum optomechanical entanglement is reached in the bistability region (for $\eta$ approaching 0) in the third regime. So the maximum achievable optomechanical entanglement is given by $\alpha$.
From Eqs .(19-22) the optimum value for effective detuning where entanglement takes its maximum value is
\begin{equation}
\Delta_{opt}=\frac{\omega_{m}}{4}\sqrt{1+\sqrt{16(\frac{\kappa}{\omega_{m}})^{2}+81}}
\end{equation}
For $\kappa=1.4\omega_{m}$ from Eq.(23) we obtain $\Delta_{opt}=0.85\omega_{m}$. Comparing this to Eq. (15)  one sees that the optimum effective detuning values for  cooling and entanglement are not the same. Even more importantly, the cooling performance is optimized for $\eta=1$, whereas entanglement becomes maximal for $\eta=0$. Using the optimum value for detuning we obtain the following expression for maximum achievable entanglement in our system
\begin{equation}
E_{N,max}=-\ln\left[\frac{1}{5}\sqrt{9+\frac{128\kappa^{2}}{8\kappa^{2}+45\omega_{m}^{2}}}\right].
\end{equation}
Note that this takes its greatest possible value for $\kappa=0$, giving $E_{N,max}=-\ln[3/5]=0.51$.

It is also interesting to look at the optomechanical entanglement for the two stable branches and their behaviour in the bistable regime. Fig. 4 shows the logarithmic negativity as a function of input power for both stable branches. Varying the input power corresponds to varying $\eta$, cf. Fig. 2.
We note the persistence of entanglement in the second stable state in a very narrow window of parameter space. As can be seen in Fig.4 the entanglement is maximum at the end of each branch, corresponding to the third regime. The fast decreasing entanglement for the second branch is in agreement with the bistability parameter behaviour in Fig. 2. Inset shows the optomechanical coupling for different stable branches. One sees clearly that the coupling constant is not the decisive parameter for the amount of entanglement in our system, and in particular that the entanglement is a non-monotonic function of the coupling constant.

\begin{figure}
\epsfig{file=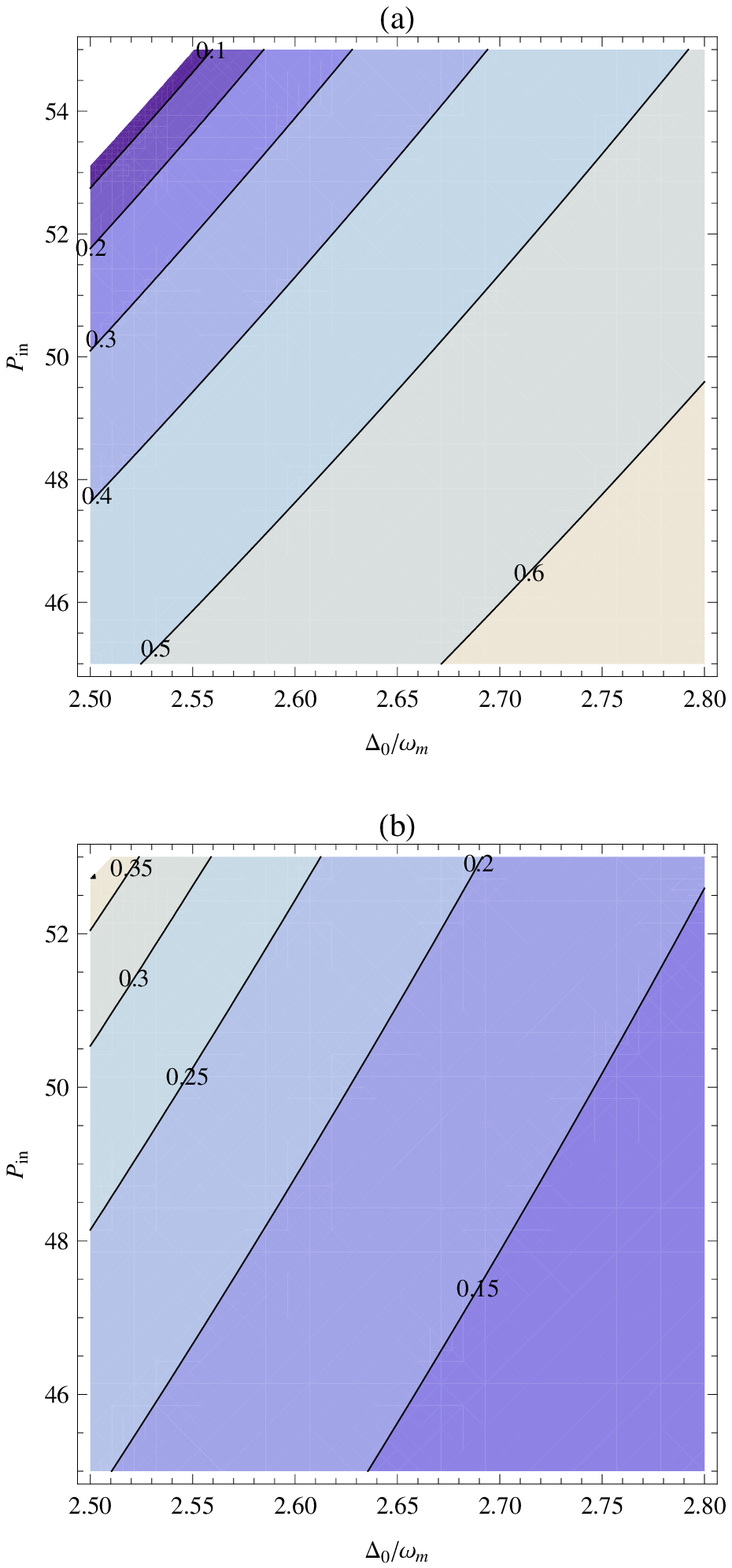,width=\columnwidth}
%\scalebox{1.7}{\includegraphics*[viewport=-3 0 350 240]{Figure5.eps}}
\caption{\label{fig3}Contour plot for bistability parameter (a) and entanglement  (b) versus bare detuning $\Delta_{0}$ and input power $P_{in}$ in $mW$. The parameters are the same as in Fig. 2}
\end{figure}

\begin{figure}[h!]
\epsfig{file=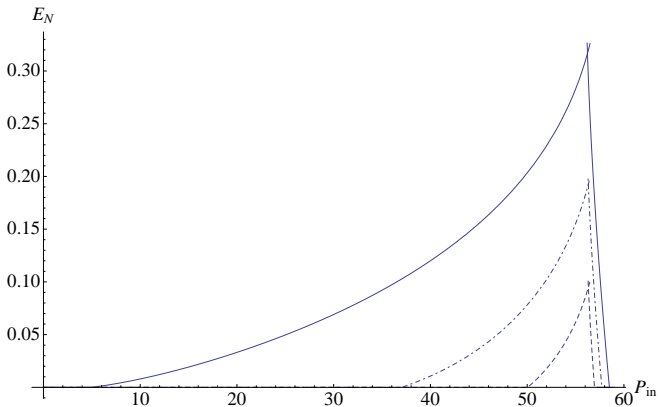,width=\columnwidth}
%\scalebox{1.8}{\includegraphics*[viewport=5 4.2 350 90]{Figure6.eps}}
\caption{\label{fig3} Plot of the logarithmic negativity versus the input power for different environment temperatures, $T=0.4K$ (solid line), $T=5$ (dot dashed), and $T=10$ (dashed). The other parameters are the same as in Fig. 2.}
\end{figure}

Until now we studied the entanglement in terms of parameters that are natural to use from a theoretical point of view. It is also interesting to look at entanglement in terms of parameters that can be directly controlled experimentally. Fig. 5  shows bistability parameter and entanglement as a function of bare detuning $\Delta_{0}$ and laser power $P$. Note that as we come close to the end of the branch for suitable detuning and large enough input power the entanglement increases.

We have also studied the robustness of entanglement with respect to the temperature. The result is shown in Fig. 6 . One sees that for higher temperatures the entanglement survives only in the vicinity of the bistable region.

Finally we note that in the recent experiment\cite{Groblacher09} the ratio of the input power to the critical power (the input power for which the bistability happen) is about 0.5. So the bistable regime should definitely be accessible experimentally.

\section{\label{Conclusion}Conclusion }

We have studied the simplest optomechanical system using the covariance matrix formalism with a special emphasis on bistability. We recovered the standard results on optomechanical cooling as a special case of our general expression for the phonon number. However, our focus was on entanglement. We identified two key parameters, namely the effective detuning and the bistability parameter (i.e. the distance from the end of each stable branch in the bistable regime), and we showed that there are different regimes for entanglement as a function of these parameters. In particular we showed that maximum entanglement is achieved when the system is simultaneously close to the red sideband (in terms of effective detuning) and close to the end of each stable branch (bistability parameter close to zero). We also showed that the dependence of entanglement on the optomechanical coupling is counter-intuitive, and that in the bistable regime the entanglement is particularly robust with respect to temperature increases.

It would be very interesting to see experimental explorations of the phenomena described in the present paper. However, it should be noted that measuring the covariance matrix, which lies at the heart of our analysis, requires direct access to the position and momentum variables of the mirror, not just the quadratures of the light. It seems that this would require either an auxiliary measurement cavity, as proposed in Ref. \cite{Vitali07}, or at least an additional laser beam. A detailed analysis of the resulting more complex dynamics is work for the future. In the present paper we focused on the entanglement characteristics of the basic system, which are already quite rich and intriguing.

{\it Acknowledgments} We thank D. Bouwmeester, A. D'Souza, D. Kleckner, A. Lvovsky and B. Pepper for very useful discussions. This work was supported by AITF and NSERC.

\end{document}